\documentclass[letterpaper]{article}

\usepackage{geometry}
\usepackage{setspace}
\usepackage{graphicx}
\usepackage{epstopdf}
\usepackage{amsmath}
\usepackage{amssymb}
\usepackage{newtxtext}
\usepackage{newtxmath}
\usepackage{url}
\usepackage{authblk}
\usepackage[style=chem-acs,sorting=none,articletitle=true,doi=false,url=false,maxnames=15]{biblatex}
\usepackage[colorlinks=true,linkcolor=blue,citecolor=blue,urlcolor=blue]{hyperref}
\DeclareCiteCommand{\cite}[\mkbibsuperscript]
  {\unskip\usebibmacro{cite:init}\usebibmacro{prenote}}
  {\usebibmacro{citeindex}\usebibmacro{cite:comp}}
  {}
  {\usebibmacro{cite:dump}\usebibmacro{postnote}}

\DeclareMathAlphabet{\mathnewtx}{OML}{ntxmi}{m}{it}
\newcommand{\wvar}{\ensuremath{\mathnewtx{w}}}
\newcommand{\xvar}{\ensuremath{\mathnewtx{x}}}

\DeclareFieldFormat{labelnumberwidth}{\mkbibparens{#1}}
\defbibenvironment{bibliography}
  {\list{}
     {\setlength{\leftmargin}{0pt}%
      \setlength{\itemindent}{0pt}%
      \setlength{\labelwidth}{0pt}%
      \setlength{\labelsep}{0pt}%
      \setlength{\itemsep}{0pt}%
      \setlength{\parsep}{0pt}%
      \setlength{\topsep}{0pt}}}
  {\endlist}
  {\item\printtext[labelnumberwidth]{\printfield{labelnumber}}\space}
\AtBeginBibliography{\setstretch{1.02}\small}

\title{Time-Reversal Invariant Topological Photonic Alloy}

\author[1]{Ruili Feng}

\author[1]{Tiantao Qu}

\author[1]{Xianbin Wu}

\author[1]{Xiaoxuan Shi}

\author[2]{Mujun He}

\author[3,4]{Lei Zhang*}

\author[1,4]{Jun Chen*}
\affil[1]{State Key Laboratory of Quantum Optics Technologies and Devices, Institute of Theoretical Physics, Shanxi University, Taiyuan 030006, China}
\affil[2]{Sanli Honors College, Shanxi University, Taiyuan 030006, China}
\affil[3]{State Key Laboratory of Quantum Optics Technologies and Devices, Institute of Laser Spectroscopy, Shanxi University, Taiyuan 030006, China}
\affil[4]{Collaborative Innovation Center of Extreme Optics, Shanxi University, Taiyuan 030006, China}

\date{*Email: \href{mailto:zhanglei@sxu.edu.cn}{zhanglei@sxu.edu.cn}; \href{mailto:chenjun@sxu.edu.cn}{chenjun@sxu.edu.cn}}

\begin{document}

\makeatletter
\twocolumn[
\begin{@twocolumnfalse}
\maketitle

\begin{abstract}
Conventional approaches generate quantum spin Hall (QSH) and quantum valley Hall (QVH) effects by breaking spatial inversion symmetry across the entire photonic crystal. Consequently, modulating the bulk transmission gaps within these periodic frameworks typically requires global modification of structural parameters throughout the lattice, severely limiting device reconfigurability and post-fabrication adaptability. In this work, we demonstrate that global inversion symmetry breaking is not a prerequisite for inducing topological phases. Instead, local and minimal symmetry breaking is sufficient. Using a photonic alloy platform based on random substitutional disorder in a parallel-plate waveguide, we show that a disorder-driven local breaking of $z$-direction mirror symmetry and a local breaking of in-plane symmetry are capable of opening topological gaps to trigger QSH and QVH phase transitions, respectively, thereby supporting robust helical edge states and valley kink states. Rigorous topological characterization via the reflection-phase winding method reveals a striking size-dependent scaling behavior in both configurations, where the threshold doping concentration required to trigger the topological phase transition asymptotically vanishes in the thermodynamic limit. By enabling flexible, on-demand bulk gap engineering simply through random doping tuning rather than global structural reconfigurations, this photonic alloy platform not only deepens understanding of disorder induced topological physics but also offers a scalable, highly efficient design strategy for integrated topological optical communications.
\end{abstract}

\section*{Keywords}
topological photonic alloys, time-reversal invariant systems, quantum spin Hall effect, quantum valley Hall effect, disorder

\vspace{1em}
\end{@twocolumnfalse}
]
\makeatother

\section{Introduction}

Over the past two decades, topological photonics has revolutionized our ability to explore exotic states of matter and control the flow of light with unprecedented robustness \cite{Lu2014,Ozawa2019TopologicalPhotonics,KhanikaevShvets2017TwoDimTopologicalPhotonics}. Whether by breaking time-reversal symmetry using gyromagnetic materials under external magnetic fields \cite{PhysRevLett.100.013904,PhysRevLett.100.013905,Wang2009,PhysRevLett.106.093903,PhysRevLett.115.253901}, or by breaking spatial inversion symmetry through lattice geometric engineering \cite{Khanikaev2013,PhysRevLett.114.127401,Cheng2016,PhysRevLett.114.223901,Xu:16,Dong2017,PhysRevB.96.020202,Gao2018NatPhys,PhysRevLett.120.063902,https://doi.org/10.1002/lpor.202300515,https://doi.org/10.1002/lpor.202503189,Xue2019SpinValley}, diverse topological phases have been successfully realized within optical platforms---most notably manifested as the photonic quantum Hall \cite{PhysRevLett.100.013904,PhysRevLett.100.013905,Wang2009,PhysRevLett.106.093903,PhysRevLett.115.253901}, quantum spin Hall (QSH) \cite{Khanikaev2013,PhysRevLett.114.127401,Cheng2016,PhysRevLett.114.223901,Xu:16}, and quantum valley Hall (QVH) effects \cite{Dong2017,PhysRevB.96.020202,Gao2018NatPhys,PhysRevLett.120.063902,https://doi.org/10.1002/lpor.202300515,https://doi.org/10.1002/lpor.202503189}.

Conventionally, weak disorder---such as edge defects created by missing unit cells or localized obstacles---is widely utilized to demonstrate the robust propagation of topological edge states \cite{Rechtsman2013PhotonicFloquet,Hafezi2013ImagingTopologicalEdgeStates,Bandres2018TopologicalLaserExperiments,Shalaev2019RobustTransport,Barik2018Science,Yang2020,https://doi.org/10.1002/lpor.201900087}. However, the role of disorder extends far beyond whether edge modes survive local perturbations. For instance, structural disorder or on-site disorder potentials can drive a transition from a trivial insulator to a topological Anderson insulator \cite{Li2009TopologicalAnderson,Groth2009TheoryTAI,PhysRevB.84.035110,Stutzer2018,PhysRevLett.125.133603,PhysRevLett.129.043902,PhysRevB.106.195304,PhysRevLett.133.133802}, whereas in amorphous systems, enhanced structural disorder can trigger a topological phase transition that reshapes the lattice order from a glass-like to a liquid-like state \cite{Zhou2020,PhysRevLett.128.056401,PhysRevB.108.L081110}. In fact, recent pioneering studies in photonic alloys \cite{PhysRevLett.132.223802} suggest that disorder can actively serve as a continuously tunable degree of freedom to manipulate topological states. Simply by continuously adjusting the random doping ratio, time-reversal-symmetry-broken photonic alloys have been shown to facilitate diverse phenomena, including band-alignment effects \cite{PhysRevB.110.094206}, transitions between first- and second-order topological phases \cite{l4hs-hb51}, the disruption of robust chiral edge propagation by weak-disorder-induced defect states despite preserved global topological invariants \cite{Shi2026Fragility}, as well as topological phase transitions in quasicrystalline \cite{10.1063/5.0232244} and amorphous alloys \cite{gw3w-5h94}, and analogous extensions to phononic platforms \cite{Li2026TopologicalAcousticAlloys}.

In this work, we focus on reciprocal photonic systems that preserve time-reversal symmetry, particularly QSH and QVH systems. Because topological phase transitions in these systems do not rely on external magnetic fields, they hold promise for scalable topological transport to optical wavelengths and enabling high-density on-chip integration. Nevertheless, conventional approaches generate QSH and QVH phases by breaking spatial inversion symmetry across the entire photonic crystal. Specifically, converting a trivial lattice into a nontrivial QSH or QVH phase typically requires the global breaking of $z$-direction mirror symmetry to introduce spin-orbit coupling \cite{Khanikaev2013,PhysRevLett.114.127401,Cheng2016}, or the global breaking of in-plane ($C_{3v}$) inversion symmetry to open a valley gap \cite{Wang2023,Lu2017,Ma_2016}. As a result, modulating the bulk transmission gaps in these periodic photonic crystal frameworks usually implements a global modification of structural parameters across the entire lattice, severely limiting device reconfigurability and post-fabrication adaptability. This raises a fundamental question: is global inversion symmetry breaking a strict prerequisite for generating QSH and QVH phases, or can localized, and even minimal, inversion symmetry breaking be sufficient to trigger these topological phase transitions?

Topological photonic alloys offer a direct route to answer this question. We demonstrate that global symmetry breaking is not a strict prerequisite for inducing topological phases. Instead, breaking inversion symmetry locally and minimally within a time-reversal-invariant photonic alloy is entirely sufficient. Utilizing a photonic alloy platform based on random substitutional disorder in a parallel-plate waveguide, we systematically explore two distinct routes: disorder-driven local breaking of $z$-direction mirror symmetry to generate helical QSH states, and local breaking of $C_{3v}$ symmetry to produce QVH kink states. Verified by reflection-phase winding analysis, both systems exhibit a striking size-dependent behavior, where the threshold doping concentration for the topological phase transition asymptotically approaches zero in the thermodynamic limit. Beyond its fundamental interest in disorder induced topological physics, this platform offers a scalable, highly efficient route for flexible gap engineering simply by adjusting the doping ratio rather than reconfiguring the entire lattice architecture.

\begin{figure*}[!t]
\centering
\includegraphics[width=1\textwidth]{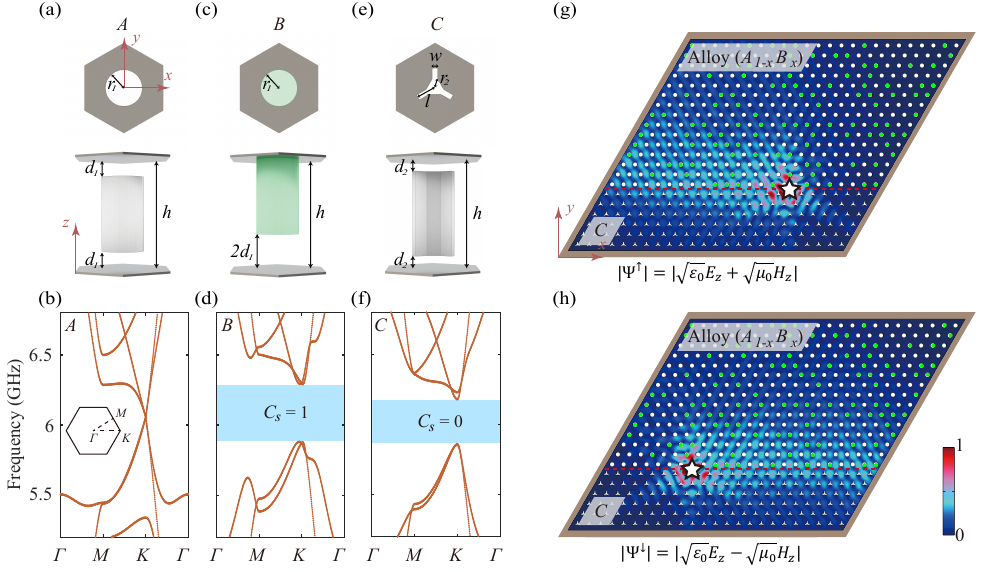}
\caption{Photonic quantum spin Hall alloy and the field distributions of its helical edge modes. (a), (c), and (e) Schematics of the $A$-type, $B$-type, and $C$-type components, respectively. The structural parameters are: lattice constant $a=36.8~\mathrm{mm}$, height $h=36.8~\mathrm{mm}$, inner radii $r_1=6.44~\mathrm{mm}$ and $r_2=3.65~\mathrm{mm}$, arm length $l=7.9~\mathrm{mm}$, arm width $\wvar=2.2~\mathrm{mm}$, and vertical displacements $d_1=2.76~\mathrm{mm}$ and $d_2=1.1~\mathrm{mm}$. (b), (d), and (f) Band structures of photonic crystals composed entirely of unit cells $A$, $B$, and $C$, respectively. The fourfold-degenerate Dirac point in (b) occurs at a frequency of $6.055~\mathrm{GHz}$. The inset in (b) depicts the first Brillouin zone, and the blue-shaded regions in (d) and (f) indicate the bandgaps. (g), (h) Simulated field distributions of the (g) spin-up ($|\Psi^{\uparrow}|$) and (h) spin-down ($|\Psi^{\downarrow}|$) modes calculated at $6.055~\mathrm{GHz}$ along the interface (red dashed line) between the $A_{1-\xvar}B_{\xvar}$ alloy ($\xvar=0.3$) and the spin-topologically trivial boundary (a photonic crystal formed by unit cell $C$). The color scale denotes the field amplitudes, and the white star indicates the position of the dipole source. The outermost boundary represented by the brown line is set to an absorbing boundary condition.}
\label{fig:spin_cells}
\end{figure*}

\section{Model and results}
In reciprocal photonic systems where time-reversal symmetry is preserved, opening a topological gap to secure non-zero topological invariants necessarily requires spatial symmetry engineering. This is typically achieved either by breaking the $z$-direction mirror symmetry to introduce spin-orbit coupling \cite{Khanikaev2013,PhysRevLett.114.127401,Cheng2016} or by breaking the in-plane ($xy$ plane) inversion symmetry \cite{PhysRevB.96.020202,Gao2018NatPhys,PhysRevLett.120.063902}. In what follows, we use these two approaches as representative examples to demonstrate, via the platform of photonic alloys \cite{PhysRevLett.132.223802,gw3w-5h94,l4hs-hb51}, that global inversion symmetry breaking is not a strict prerequisite for generating topological edge states. In fact, local inversion-symmetry breaking---which can asymptotically vanish in the thermodynamic limit of large system sizes---suffices. Furthermore, the strength of this local symmetry breaking, governed by the alloy's doping concentration, provides an effective knob for dynamically tuning the topological gap.
\subsection{Quantum spin Hall photonic alloys}
Photons are bosons and therefore inherently lack the Kramers degeneracy that arises directly from time-reversal symmetry in electronic systems \cite{RevModPhys.82.3045,RevModPhys.83.1057,PhysRevLett.95.226801}. Consequently, realizing the quantum spin Hall effect (QSHE) in a reciprocal photonic system necessitates leveraging specific spatial lattice symmetries and electromagnetic mode hybridizations to construct a pair of degenerate states that emulate spin-up ($\Psi^{\uparrow}$) and spin-down ($\Psi^{\downarrow}$), such that the effective time-reversal operator satisfies $\hat{T}_{\mathrm{eff}}^2=-1$ \cite{Cheng2016}. Therefore, conventional photonic QSHE implementations are typically based on lattice structures with high spatial symmetry, such as the honeycomb lattice formed by unit cell $A$ shown in Figure~\ref{fig:spin_cells}a. By finely tuning the geometric parameters of the photonic crystal unit cell, a degeneracy between the TE and TM bands can be induced at a high-symmetry point in momentum space \cite{Khanikaev2013,PhysRevLett.114.127401,PhysRevB.95.165102}. As shown in Figure~\ref{fig:spin_cells}b, this manifests as a fourfold-degenerate double Dirac cone at the $K$ point. At this degeneracy, the modes mathematically span a perfect two-dimensional Hilbert subspace, enabling their linear combinations to define stable, orthogonal spin-up and spin-down states \cite{Khanikaev2013,PhysRevLett.133.133802}:
\begin{equation}
|\Psi^{\uparrow}|=|\sqrt{\varepsilon_0}E_z+\sqrt{\mu_0}H_z|,\quad
|\Psi^{\downarrow}|=|\sqrt{\varepsilon_0}E_z-\sqrt{\mu_0}H_z| .
\label{eq:pseudospin}
\end{equation}
To lift the degeneracy and open a topological bandgap, it is necessary to break the mirror symmetry along the $z$-direction to introduce spin-orbit coupling \cite{Cheng2016}. This structural asymmetry introduces off-diagonal coupling elements into the Hamiltonian within the TE-TM basis, inducing a hybridization between the TE and TM modes. Actually, upon applying a unitary transformation from the TE-TM basis to the circularly polarized pseudospin basis defined in eq~\ref{eq:pseudospin}, the resulting effective Hamiltonian remains diagonal \cite{PhysRevLett.114.127401}, leaving the $\Psi^{\uparrow}$ and $\Psi^{\downarrow}$ modes decoupled. As illustrated in Figures~\ref{fig:spin_cells}c and \ref{fig:spin_cells}d, constructing a honeycomb lattice using $B$-type unit cells opens a topological gap characterized by a spin Chern number $C_s=1$ (detailed calculations of the topological invariants are provided in the Supporting Information), thereby driving a transition from the trivial photonic crystal built from $A$-type unit cells to a QSH photonic crystal \cite{PhysRevLett.114.127401}.

Here, we will show that a local breaking of the $z$-direction mirror symmetry suffices to achieve the photonic QSHE. Specifically, we introduce random substitutional disorder by doping $B$-type components into the honeycomb lattice of unit cell $A$, thereby locally breaking the $z$-direction mirror symmetry. We construct this substitutional photonic alloy, $A_{1-\xvar}B_{\xvar}$, to investigate whether the photonic QSHE robustly persists under this condition. The doping concentration $\xvar$ is defined as \cite{PhysRevLett.132.223802}
\begin{equation}
\xvar=\frac{N_B}{N_A+N_B}.
\label{eq:concentration}
\end{equation}
Here, $N_A$ and $N_B$ denote the numbers of $A$-type and $B$-type components within the alloy region, respectively.

To clearly visualize the edge states, the alloy region is interfaced on one side with a honeycomb-lattice photonic crystal composed of $C$-type unit cells [see Figure~\ref{fig:spin_cells}e], as shown in Figures~\ref{fig:spin_cells}g and \ref{fig:spin_cells}h. The $C$-type photonic crystal \cite{PhysRevB.95.165102,PhysRevLett.133.133802} is chosen as the boundary because it supports propagation of both TM and TE modes while exhibiting a spin-topologically trivial bandgap [spin Chern number $C_s=0$, as shown in Figure~\ref{fig:spin_cells}f] within a closely overlapping frequency range. This configuration effectively blocks bulk mode propagation, thereby confining the topological photons along the interface and preventing radiative leakage into the bulk. Furthermore, maintaining lattice continuity across the interface preserves the local spatial symmetries, ensuring that the effective time-reversal symmetry ($\hat{T}_{\mathrm{eff}}$) does not degrade at the boundary and thus safeguarding the clean spin-momentum locking \cite{Khanikaev2013,Cheng2016,Kang2018} of the emergent edge states.

From the field distributions of $\Psi^{\uparrow}$ and $\Psi^{\downarrow}$ shown in Figures~\ref{fig:spin_cells}g and \ref{fig:spin_cells}h (obtained using COMSOL Multiphysics with the finite-element method), it is evident that edge states clearly emerge even at a low doping level of $\xvar=0.3$. The fields remain localized along the boundary, with $\Psi^{\uparrow}$ and $\Psi^{\downarrow}$ displaying the signature counter-propagating characteristics dictated by spin-momentum locking~\cite{Khanikaev2013,PhysRevLett.114.127401,Cheng2016,Kang2018}. Next, focusing on the $\Psi^{\uparrow}$ mode as a representative example, we rigorously demonstrate that these edge states emerging within the photonic alloy are indeed topologically protected helical edge modes.

\begin{figure*}[!t]
\centering
\includegraphics[width=1\textwidth]{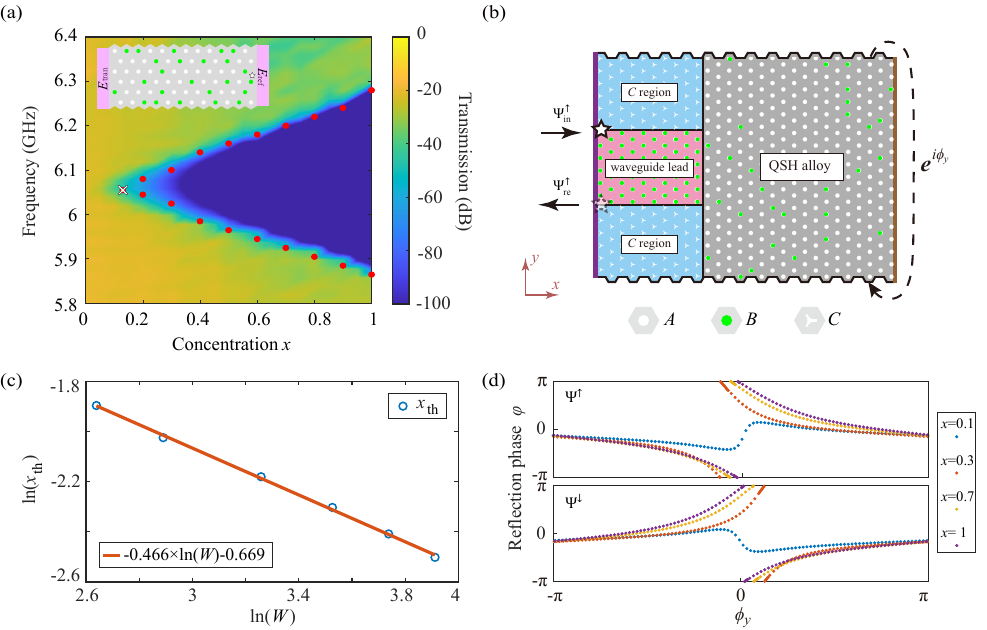}
\caption{Topological properties of photonic alloys. (a) Bulk transmission spectrum as a function of the doping concentration $\xvar$. Each data point represents the statistical average over 5 random configurations. The red dots indicate the boundary of the topological gap characterized by a spin Chern number of $C_s=1$. For each doping concentration, the gap boundary is determined via the reflection phase winding method by analyzing 5 random configurations. Inset: Schematic configuration for the numerical transmission calculation in the $A_{1-\xvar}B_{\xvar}$ photonic alloy system, where white and green dots denote $A$-type and $B$-type components, respectively, and the white star represents the excitation dipole source. (b) Schematic setup for the reflection phase winding calculation, consisting of a square photonic alloy domain ($W \times W$), a waveguide lead composed of unit cells $B$, and a spin-topologically trivial interface constructed from unit cell $C$. The leftmost boundary (purple) of the system is set as an impedance boundary condition, the rightmost boundary (brown) as an absorbing boundary condition, and the top and bottom boundaries are connected via a twisted boundary condition. The white solid and white dashed stars mark the respective dipole source locations when probing the topological properties of the $\Psi^{\uparrow}$ and $\Psi^{\downarrow}$ modes (e.g., an incident wave $\Psi_{\rm in}^{\uparrow}$ enters the system, and the corresponding reflected mode $\Psi_{\rm re}^{\uparrow}$ is analyzed). (c) Sample size dependence of the threshold doping concentration $\xvar_{\rm th}$ at $6.055~\mathrm{GHz}$. Each data point (blue open circles) is computed with 5 random configurations. The fitted curve (red line) indicates that the threshold doping concentration for the topological phase transition asymptotically approaches zero as the system size increases. (d) Reflection phase $\varphi$ as a function of the twisting angle $\phi_y$ for the two spin modes at $\xvar=0.1$, $0.3$, $0.7$, and $1$.}
\label{fig:spin_transmission}
\end{figure*}

We begin by analyzing the bulk transmission spectrum~\cite{PhysRevLett.132.223802} of the photonic alloy ($A_{1-\xvar}B_{\xvar}$) across various doping concentrations, as presented in Figure~\ref{fig:spin_transmission}a, to track the evolution of the transmission gap. For the bulk transmission simulations [see the inset of Figure~\ref{fig:spin_transmission}a], the top and bottom boundaries of the photonic alloy (with dimensions of $40a \times 12a$) are connected via continuous periodic boundary conditions, while absorbing boundary conditions are applied to the left and right sides. A dipole source is positioned near the right boundary. The time-averaged Poynting vector is integrated over the left and right boundaries to calculate the energy. Defining $E_{\rm tran}$ as the energy passing through the photonic alloy via the left boundary and $E_{\rm ref}$ as the energy directly reflected by the photonic alloy that exits through the right boundary, the total energy leaving the system is given by $E_{\rm tot}=E_{\rm tran}+E_{\rm ref}$. The bulk transmission is then evaluated as $\langle T\rangle=\left\langle 20\log_{10}\left(E_{\rm tran}/{E_{\rm tot}}\right)\right\rangle$, where $\langle \cdot \rangle$ denotes the statistical average over multiple disordered configurations at a fixed doping concentration. As illustrated in Figure~\ref{fig:spin_transmission}a, increasing the doping concentration $\xvar$ causes a bulk transmission gap to gradually open and broaden around the central frequency of $6.055~\mathrm{GHz}$. In conventional quantum spin Hall photonic crystals, tuning the bulk transmission gap typically requires a global reconfiguration of the structural parameters across the entire lattice, such as altering the vertical displacement $d_1$~\cite{PhysRevLett.114.127401,Cheng2016} or the ratio of the inner radius $r_1$ to the lattice constant $a$~\cite{PhysRevLett.114.223901,Xu:16}. By contrast, Figure~\ref{fig:spin_transmission}a reveals that the photonic alloy platform enables flexible control over the gap simply via the random substitution of $A$-type components with $B$-type components. This approach provides a highly convenient and efficient route for gap engineering.

To prove that this opened bulk transmission gap possesses a non-trivial topological character, we further employ the reflection phase winding method~\cite{PhysRevLett.129.043902,Li2026TopologicalAcousticAlloys} to characterize the topology of the photonic alloy. Specifically, the photonic alloy ($A_{1-\xvar}B_{\xvar}$) is connected to a waveguide lead formed by unit cell $B$ and then interfaced with a spin-topologically trivial photonic crystal constructed from unit cell $C$, as schematically detailed in Figure~\ref{fig:spin_transmission}b. When probing the topological properties of the $\Psi^{\uparrow}$ and $\Psi^{\downarrow}$ modes, the corresponding dipole source locations are marked by a white solid star and a white dashed star, respectively. The leftmost boundary of the system is set as an impedance boundary to facilitate a well-defined tracking of the reflection phase, the rightmost boundary is set as an absorbing boundary, and the top ($y=W$) and bottom ($y=0$) boundaries are linked via a twisted boundary condition~\cite{doi:10.1126/sciadv.adg3186}:
\begin{equation}
\Psi^\sigma(y=W)=e^{i\phi_y}\Psi^\sigma(y=0),\quad \sigma=\uparrow,\downarrow.
\label{eq:twisted_boundary}
\end{equation}
This twisted boundary condition can be viewed as an adiabatically varying gauge flux $\phi_y$ threading through the hollow of the rolled-up photonic system \cite{PhysRevB.85.165409,PhysRevB.83.155429}. In analogy with the Wilson-loop formalism~\cite{Wang_2019}---where the winding number of the Berry phase along one reciprocal lattice basis vector corresponds to the Chern number in periodic systems~\cite{doi:10.1143/JPSJ.74.1674,Wang_2019}---the Chern number in disordered photonic systems is given by the winding number of the reflection phase $\varphi$ (defined as the phase difference between the incident and reflected waveguide modes) as the twisting angle $\phi_y$ varies from $-\pi$ to $\pi$ over one full modulation cycle. At the excitation frequency of $f=6.055~\mathrm{GHz}$, we examine the dependence of the reflection phase $\varphi$ on the twisting angle $\phi_y$ across several doping concentrations, as presented in Figure~\ref{fig:spin_transmission}d. As expected, for the QSH photonic crystal at a doping concentration of $\xvar=1$ (composed entirely of unit cell $B$), the reflection phase of the $\Psi^{\uparrow}$ mode exhibits a full $2\pi$ winding, signaling a nontrivial Chern number of $C_{\uparrow}=1$. Concurrently, the reflection phase of the $\Psi^{\downarrow}$ mode exhibits a full $-2\pi$ winding, signaling a nontrivial Chern number of $C_{\downarrow}=-1$. This yields a total spin Chern number of $C_s=(C_{\uparrow}-C_{\downarrow})/2=1$~\cite{Khanikaev2013,PhysRevLett.114.127401,Cheng2016}, consistent with the value obtained using the Wilson-loop approach [see Supporting Information and Figure~\ref{fig:spin_cells}d]. For the intermediate photonic alloy cases with $\xvar=0.3$ and $\xvar=0.7$ [which lie within the bulk transmission gap in Figure~\ref{fig:spin_transmission}a], the reflection phases of both the $\Psi^{\uparrow}$ and $\Psi^{\downarrow}$ modes display the same full $2\pi$ and $-2\pi$ windings observed in the QSH photonic crystal. This winding yields nontrivial Chern numbers of $C_{\uparrow}=1$ and $C_{\downarrow}=-1$, thereby demonstrating that the photonic alloy possesses a spin Chern number of $C_s=1$, and that its corresponding $\Psi^{\uparrow}$ and $\Psi^{\downarrow}$ modes are topologically protected helical edge states. Conversely, at a lower doping concentration of $\xvar=0.1$, the reflection phase for both spin modes fails to achieve a full $\pm 2\pi$ winding, corresponding to a topologically trivial phase. Utilizing this reflection phase winding method, we mark [with red dots in Figure~\ref{fig:spin_transmission}a] the boundary of the topological gap characterized by a spin Chern number of $1$, confirming that the transmission gap opened via disorder is indeed the topological gap.

Notably, the threshold doping concentration $\xvar_{\rm th}$ marking the topological phase transition from a trivial medium to a nontrivial QSH photonic alloy [indicated by the cross in Figure~\ref{fig:spin_transmission}a] asymptotically approaches zero as the system size increases, a trend clearly illustrated by the scaling behavior of the threshold concentration in Figure~\ref{fig:spin_transmission}c. At the excitation frequency of $f=6.055~\mathrm{GHz}$, the threshold doping concentration plotted against the system size exhibits an approximately linear relationship on a log--log scale, well described by the empirical relation
\begin{equation}
\ln \xvar_{\rm th}=-0.466\ln W-0.669,
\label{eq:scaling}
\end{equation}
namely $\xvar_{\rm th}\propto W^{-0.466}$. This implies that the threshold doping concentration of the substitutional photonic alloy indeed approaches zero in the thermodynamic limit. Consequently, within a sufficiently large sample, even a minute concentration of $B$-type components is capable of driving the system into a topologically nontrivial QSH phase. This result can be readily understood from the band structures in Figure~\ref{fig:spin_cells}b. The photonic crystal formed by unit cell $A$ features a gapless Dirac point at the $K$ point at $6.055~\mathrm{GHz}$, making its degeneracy highly susceptible to external perturbations. Since this degenerate frequency falls within the bulk topological bandgap of the unit cell $B$ crystal [Figure~\ref{fig:spin_cells}d], a small amount of $B$-type doping is sufficient to open a topological gap.

\begin{figure*}[!t]
\centering
\includegraphics[width=1\textwidth]{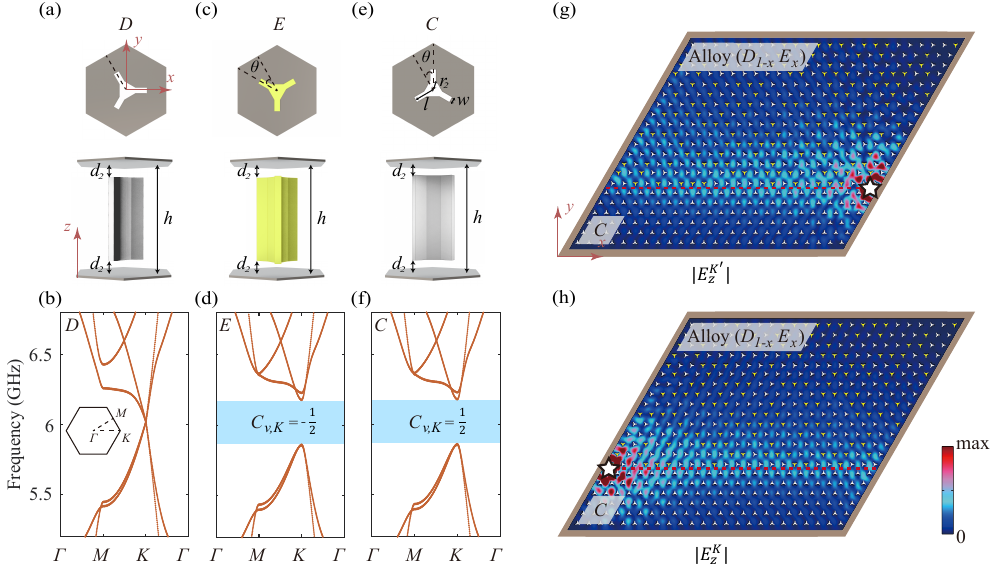}
\caption{Topological valley Hall photonic alloy and the field distributions of its valley kink modes. (a), (c), and (e) Schematics of the $D$-type, $E$-type, and $C$-type components, respectively. The $E$-type and $C$-type configurations are obtained by rotating the internal tripod structure of the $D$-type component counterclockwise and clockwise by $\theta = 30^\circ$. (b), (d), and (f) Band structures of photonic crystals composed entirely of unit cells $D$, $E$, and $C$, respectively. The gapless Dirac point of the $D$-type photonic crystal appears at a frequency of $6.02~\mathrm{GHz}$. The inset in (b) depicts the first Brillouin zone, and the blue-shaded regions in (d) and (f) indicate the bandgaps. (g), (h) Simulated field distributions along the interface (indicated by the red dashed lines) between the $\text{D}_{1-\xvar}\text{E}_\xvar$ alloy ($\xvar=0.4$) and the $C$-type photonic crystal under selective excitation of the (g) $K'$ valley ($|E_z^{K'}|$) and (h) $K$ valley ($|E_z^K|$) at $6.02~\mathrm{GHz}$. The color scale denotes the field amplitudes, the white star marks the position of the dipole source, and the outermost brown line represents the absorbing boundary condition.}
\label{fig:valley_cells}
\end{figure*}

\subsection{Quantum valley Hall photonic alloys}

Having shown that breaking the local $z$-direction mirror symmetry can successfully induce a QSH phase, we now extend this alloy strategy to a different time-reversal-invariant system. Specifically, we investigate whether breaking only a minimal, local in-plane spatial inversion symmetry is sufficient to trigger the quantum valley Hall effect (QVHE). To illustrate this mechanism, we examine a substitutional photonic alloy system, $D_{1-\xvar}E_{\xvar}$, created by randomly substituting $E$-type components [Figure~\ref{fig:valley_cells}c] into the honeycomb lattice of unit cell $D$ [Figure~\ref{fig:valley_cells}a], where the doping concentration $\xvar$ is given by
\begin{equation}
\xvar=\frac{N_E}{N_D+N_E}.
\label{eq:qvh_concentration}
\end{equation}
Here, $N_D$ and $N_E$ denote the numbers of $D$-type and $E$-type structural components within the alloy region, respectively.

In a honeycomb-lattice photonic crystal composed solely of unit cell $D$ [$x=0$, see Figure~\ref{fig:valley_cells}a], although the spatial inversion symmetry is broken, the system still preserves $C_{3v}$ symmetry. Under the combined protection of time-reversal invariance and vertical mirror symmetry ($\sigma_v$)~\cite{PhysRevB.89.134302,He2015DiracMirror,Makwana2019}, the Dirac cones remain symmetry-protected. Consequently, the band structures shown in Figure~\ref{fig:valley_cells}b exhibit a gapless Dirac cone pinned at $6.02~\mathrm{GHz}$ at the $K$ point. In conventional valley photonic crystals~\cite{PhysRevB.95.165102,Gao2018NatPhys}, a widely used approach to lift this valley degeneracy and open a bandgap is to reduce the symmetry from $C_{3v}$ to $C_3$. As illustrated in Figures~\ref{fig:valley_cells}c and \ref{fig:valley_cells}e, this is achieved by rotating the internal tripod structure by $\pm 30^\circ$, which explicitly breaks the vertical mirror symmetry $\sigma_v$. Counter-rotations open up valley bandgaps with inverted Berry curvature distributions (see Supporting Information), yielding opposite valley Chern numbers, as shown in Figures~\ref{fig:valley_cells}d and \ref{fig:valley_cells}f. While such a global symmetry reduction is a well-established route to induce the QVHE, it remains to be seen whether a topological phase transition from a trivial state to the QVH phase can be triggered by breaking the $C_{3v}$ symmetry only locally and minimally.

To address this, we introduce a small amount of randomly distributed $E$-type components into the honeycomb-lattice photonic crystal composed of unit cell $D$ to break the local $C_{3v}$ symmetry. To explore the topological edge transport of photons, we adopt a widely used approach for probing valley edge states by interfacing one side of the photonic alloy $D_{1-\xvar}E_{\xvar}$ with a valley photonic crystal formed by unit cell $C$, as illustrated in Figures~\ref{fig:valley_cells}g and \ref{fig:valley_cells}h. Because the periodic $E$-type and $C$-type photonic crystals carry opposite valley Chern numbers at the $K$ valley, i.e., $C_{v,K}^{E}=-1/2$ and $C_{v,K}^{C}=1/2$, a non-zero integer jumping of the valley Chern number ($\Delta C_{v,K}$) is established across the interface between the $E$-type and $C$-type domains
\begin{equation}
\begin{aligned}
\Delta C_{v,K} &= C_{v,K}^{E}-C_{v,K}^{C}=-1.
\end{aligned}
\label{eq:valley_interface_index}
\end{equation}
This integer difference determines both the number and the propagation direction of the valley kink states~\cite{Gao2018NatPhys} at the interface. Furthermore, since the Berry curvature is an odd function of the wave vector $\mathbf{k}$ under time-reversal symmetry, integrating the Berry curvature around the $K'$ valley yields a valley Chern number opposite to that at $K$, leading to $\Delta C_{v,K'}=1$. These constraints dictate that the valley kink states exhibit counter-propagating behavior when selectively excited at the $K$ and $K'$ valleys. Once the photonic alloy $D_{1-\xvar}E_{\xvar}$ develops the same valley-Hall character as the pure $E$-type photonic crystal, this distinctive valley-momentum locking feature is expected to emerge at the interface. Remarkably, as observed in Figures~\ref{fig:valley_cells}g and \ref{fig:valley_cells}h, at a doping concentration of $\xvar=0.4$, the interface (indicated by red dashed lines) between the photonic alloy $D_{1-\xvar}E_{\xvar}$ and the $C$-type domain clearly exhibits optical fields that are confined along the boundary and propagate in opposite directions under selective excitation of the $K$ and $K'$ valleys. Notably, these propagation directions match perfectly with those dictated by $\Delta C_{v,K}=-1$ and $\Delta C_{v,K'}=1$. Next, we demonstrate that these edge modes supported at the $D_{1-\xvar}E_{\xvar}/C$ interface are indeed topologically protected valley kink states.

\begin{figure}[!htbp]
\centering
\includegraphics[width=\columnwidth]{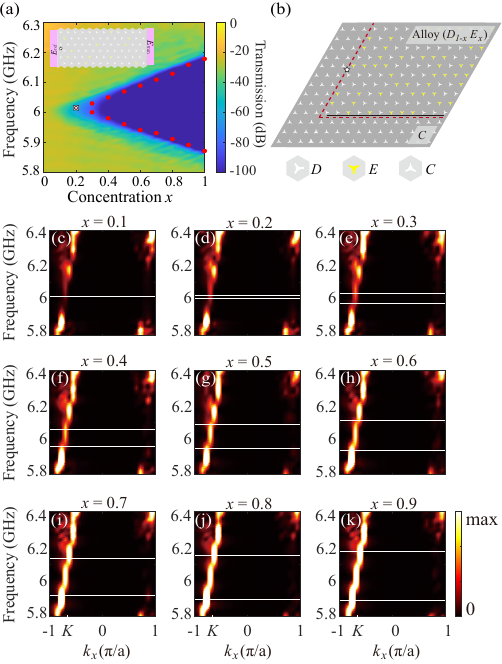}
\caption{Evolution of the bulk transmission gap and momentum-space edge modes for the QVH photonic alloy. (a) Bulk transmission spectrum as a function of the doping concentration $\xvar$ for the $D_{1-\xvar}E_{\xvar}$ photonic alloy. Each data point is obtained by averaging over five random configurations, with the red dots tracking the boundary of the topological gap. Inset: Schematic configuration for the numerical transmission calculation with an alloy domain of dimensions $40a \times 12a$, where white and yellow elements denote $D$-type and $E$-type components, respectively. (b) Configuration for extracting the real-space electric field along the boundary. The red dashed line denotes the interface between the $D_{1-\xvar}E_{\xvar}$ alloy and the $C$-type photonic crystal, while the black solid line (of length $26a$) indicates the path along which the electric field data is extracted for the spatial Fourier transform. The white star marks the position of the excitation dipole source in (a) and (b). (c)--(k) Evolution of the edge-mode profile in momentum space for the photonic alloy $D_{1-\xvar}E_{\xvar}$ across doping concentrations from $\xvar=0.1$ to $0.9$. The color scale denotes the Fourier amplitude of the extracted electric fields. The white reference line in (c) marks the Dirac point frequency of the $D$-type photonic crystal, whereas the white reference lines in (d)--(k) mark the lower and upper boundaries of the transmission gap extracted from (a) at the corresponding doping concentrations.}
\label{fig:valley_fourier}
\end{figure}

As exemplified by the excitation at the $K$ valley, the bulk transmission spectrum of the photonic alloy $D_{1-\xvar}E_{\xvar}$ in Figure~\ref{fig:valley_fourier}a shows that a bulk transmission gap opens and broadens around the Dirac point frequency ($6.02~\mathrm{GHz}$) of the $D$-type photonic crystal as the doping concentration $\xvar$ increases. To provide a clear visualization of the edge-mode profile evolution, we examine the field modes in momentum space [Figures~\ref{fig:valley_fourier}c--\ref{fig:valley_fourier}k], obtained via a Fourier transform of the real-space electric field data extracted along the boundary (black line) schematically detailed in Figure~\ref{fig:valley_fourier}b. At a low doping concentration of $\xvar=0.1$, no edge states emerge near the $6.02~\mathrm{GHz}$ frequency [indicated by the white line in Figure~\ref{fig:valley_fourier}c], which directly corresponds to the un-opened gap regime in Figure~\ref{fig:valley_fourier}a. As $\xvar$ progressively increases, the bulk transmission gap gradually widens around the center frequency of $6.02~\mathrm{GHz}$. Concurrently, as illustrated in Figures~\ref{fig:valley_fourier}d and \ref{fig:valley_fourier}e, a continuous band gradually emerges around the $K$ valley within the gap region (bounded by the two solid white lines). With further increases in $\xvar$, the gap in Figure~\ref{fig:valley_fourier}a broadens substantially, and the edge states in the gap region become more evident. These findings demonstrate a clear correlation between the formation of edge modes and the evolution of the bulk gap, confirming that the modes at the $D_{1-\xvar}E_{\xvar}/C$ interface do not originate from accidental, boundary-localized resonances. The distinct band profile near the $K$ point in Figures~\ref{fig:valley_fourier}c--\ref{fig:valley_fourier}k also indicates that, although the random substitution in the photonic alloy $D_{1-\xvar}E_{\xvar}$ introduces spatial disorder, the relatively small rotation angle ($\theta=30^\circ$) of the doped $E$-type components keeps the perturbation not strong enough to induce significant intervalley mixing~\cite{Gao2018NatPhys}. As a result, the $K$ and $K'$ valleys remain well defined, allowing the valley Hall topology to persist even within the disordered photonic alloy system.

\begin{figure}[!htbp]
\centering
\includegraphics[width=\columnwidth]{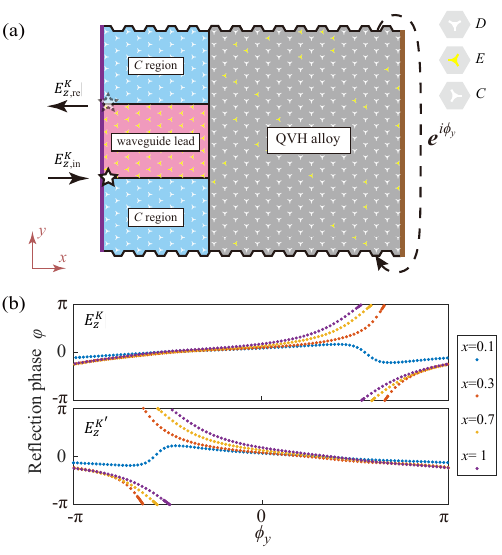}
\caption{Topological characterization of the photonic alloy via the reflection-phase winding method. (a) Schematic setup for the reflection phase winding calculation, consisting of a square photonic alloy $D_{1-\xvar}E_{\xvar}$ domain, a waveguide lead composed of unit cell $E$, and an interface constructed from unit cell $C$. The applied boundary conditions are the same as those used in the QSH reflection-phase setup. The white solid and white dashed stars mark the respective dipole source locations when probing the $K$ and $K'$ valley modes (e.g., an incident wave $E_{z,\mathrm{in}}^{K}$ enters the system, and the corresponding reflected mode $E_{z,\mathrm{re}}^{K}$ is analyzed). (b) Reflection phase $\varphi$ as a function of the twisting angle $\phi_y$ at doping concentrations of $\xvar=0.1$, $0.3$, $0.7$, and $1$ under selective $K$ and $K'$ valley excitations.}
\label{fig:valley_scattering}
\end{figure}

To rigorously identify the topology of these edge states inside the bulk transmission gap, we employ the reflection-phase winding method. As shown in Figure~\ref{fig:valley_scattering}, the setup is analogous to that used for the QSH photonic alloy in Figure~\ref{fig:spin_transmission}b, except that the incoming and reflected channels here are valley modes rather than spin modes. As detailed in the schematic of Figure~\ref{fig:valley_scattering}a, the alloy domain is connected to a waveguide lead composed of unit cell $E$ and then interfaced with a valley photonic crystal constructed from unit cell $C$. In the periodic limit ($\xvar=1$), the interface simplifies to a boundary between $E$-type and $C$-type valley photonic crystals. As illustrated in Figure~\ref{fig:valley_scattering}b, under selective excitation of the $K$ ($K'$) valley, the reflection phase $\varphi$ winds by $-2\pi$ ($2\pi$) as the twisting angle $\phi_y$ varies from $-\pi$ to $\pi$ over one full modulation cycle, precisely corresponding to the topological invariants $\Delta C_{v,K}=-1$ ($\Delta C_{v,K'}=1$). For the intermediate alloy cases, such as $\xvar=0.3$ and $\xvar=0.7$, the reflection phases under selective $K$ and $K'$ valley excitations exhibit identical, opposite full windings. This proves that the $D_{1-\xvar}E_{\xvar}$ photonic alloy acquires the same QVH topology as the pure $E$-type valley photonic crystal, confirming that the edge modes generated at the $D_{1-\xvar}E_{\xvar}/C$ interface are topologically protected valley kink states. In contrast, at the lower doping concentration of $\xvar=0.1$, the reflection phase fails to achieve a full winding under either valley excitation, signifying that the photonic alloy $D_{1-\xvar}E_{\xvar}$ remains topologically trivial in this regime. Utilizing this reflection-phase winding method, we mark the boundary of the topological gap with red dots in Figure~\ref{fig:valley_fourier}a, demonstrating that locally breaking $C_{3v}$ symmetry (via disordered doping of $E$-type components into the $D$-type photonic crystal) indeed opens a topological gap and induces the QVHE.

Furthermore, the bulk band structure of the $D$-type photonic crystal in Figure~\ref{fig:valley_cells}b highlights that the linear dispersion surrounding the gapless Dirac point at the $K$ point renders its degeneracy highly susceptible to external perturbations. Since this degenerate frequency lies within the topological bandgap of the $E$-type photonic crystal [Figure~\ref{fig:valley_cells}d], only a small amount of $E$-type doping is sufficient to lift the degeneracy and open a topological gap. Consequently, akin to the QSH photonic alloy case illustrated in Figure~\ref{fig:spin_transmission}c, the threshold doping concentration $\xvar_{\rm th}$ [indicated by the cross in Figure~\ref{fig:valley_fourier}a] marking the topological phase transition from a trivial medium to a nontrivial QVH photonic alloy is expected to asymptotically approach zero as the system size increases. Due to the substantial computational cost, we do not present the asymptotic scaling curve here. Nevertheless, these collective results definitively demonstrate that a topological phase transition from a trivial state to the QVH phase can be triggered by breaking the $C_{3v}$ symmetry only locally and minimally.

\section{Conclusion and discussion}
By employing a photonic alloy platform based on random substitutional disorder in a parallel-plate waveguide, we demonstrate that global inversion symmetry breaking is not a strict prerequisite for generating topological edge states in time-reversal-invariant systems. Instead, breaking inversion symmetry only locally and minimally is sufficient. We study two routes to inversion-symmetry breaking: one that breaks the $z$-direction mirror symmetry to introduce spin-orbit coupling, and another that breaks the in-plane ($xy$-plane) inversion symmetry.

In the former system, we discover that disordered, local breaking of the $z$-direction mirror symmetry opens a topological gap from the double Dirac cone, generating helical edge states that display the counter-propagating characteristics dictated by spin-momentum locking. The topology is verified via the reflection-phase winding method. We further find that the threshold doping concentration marking the topological phase transition from a trivial medium to a nontrivial QSH photonic alloy decreases with increasing system size, asymptotically approaching zero in the thermodynamic limit. In the latter system, we find that disorder-induced, local breaking of $C_{3v}$ symmetry opens a topological gap and produces valley kink states that control photonic transport via the valley degree of freedom. In addition to a rigorous topological characterization of the QVH photonic alloy using the reflection-phase winding method, we provide a clear visualization of the edge-mode profile evolution concurrently with the bulk gap opening. Akin to the QSH case, the threshold doping concentration for the QVH topological phase transition is also expected to asymptotically approach zero as the system size increases, definitively showing that local and minimal symmetry breaking suffices to trigger a topological phase transition.

These results also highlight the distinct practical advantages of using time-reversal-invariant photonic alloys for topological systems. Unlike conventional photonic crystals, where tuning the bulk transmission gap typically requires a global reconfiguration of structural parameters across the entire lattice, the photonic alloy platform enables flexible gap engineering simply by adjusting the random-substitution doping concentration. This approach provides a highly convenient and efficient route for gap modulation and the exploration of topological phase transitions. Moreover, because the proposed alloys achieve topological phases without relying on external magnetic fields or gyromagnetic materials, they hold promise for scaling topological transport down to optical wavelengths. Collectively, our work not only deepens the theoretical understanding of topological phase transitions in reciprocal photonic systems and the interactions between topological edge states and disorder, but also provides important guidance for designing optical communication devices and particle manipulation platforms leveraging topological edge modes.

\section*{Associated Content}
\subsection*{Supporting Information Available}
The Supporting Information is available free of charge.

Topological invariants for periodic photonic crystals (PDF)

\section*{Author Information}
\subsection*{Corresponding Author}
\begingroup
\setlength{\parindent}{0pt}
*Lei Zhang -- State Key Laboratory of Quantum Optics Technologies and Devices, Institute of Laser Spectroscopy, Shanxi University, Taiyuan 030006, China; Collaborative Innovation Center of Extreme Optics, Shanxi University, Taiyuan 030006, China.\\
Email: \href{mailto:zhanglei@sxu.edu.cn}{zhanglei@sxu.edu.cn}

*Jun Chen -- State Key Laboratory of Quantum Optics Technologies and Devices, Institute of Theoretical Physics, Shanxi University, Taiyuan 030006, China; Collaborative Innovation Center of Extreme Optics, Shanxi University, Taiyuan 030006, China.\\
Email: \href{mailto:chenjun@sxu.edu.cn}{chenjun@sxu.edu.cn}
\endgroup

\subsection*{Author Contributions}
The manuscript was written through contributions of all authors. All authors have given approval to the final version of the manuscript.

\subsection*{Funding Sources}
This work was supported by the Quantum Science and Technology-National Science and Technology Major Project (Grant No. 2025ZD0300200), the National Natural Science Foundation of China (Grants No. 12574340 and No. 12474047), the Fund for Shanxi 1331 Project, and the research project supported by Shanxi Scholarship Council of China.

\subsection*{Data Availability Statement}
The data that support the findings of this article are not publicly available. The data are available from the authors upon reasonable request.

\subsection*{Notes}
The authors declare no competing financial interest.

\printbibliography

\clearpage
\onecolumn
\begin{center}
{\large Table of Contents artwork\par}
\vspace{0.6cm}
\includegraphics[width=3.25in,height=1.75in,keepaspectratio]{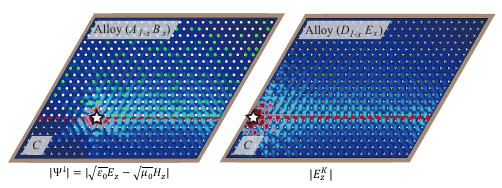}
\vspace{0.6cm}

\noindent Random local symmetry breaking in time-reversal-invariant photonic alloys opens tunable topological gaps and supports robust helical and valley kink transport.
\end{center}

\end{document}